\documentclass[floatfix,aps,twocolumn,showpacs,superscriptaddress]{revtex4-1}
\usepackage{amsmath}
\usepackage{amssymb}
\usepackage{graphicx}
\usepackage{dcolumn}
\usepackage{natbib}

\newcommand{\mean}[1]{\left\langle {#1} \right\rangle }
\newcommand{\p}[0]{p_\mathrm{co}}
\newcommand{\hq}[0]{{\hat Q}_6}
\newcommand{\hc}[0]{\hat{C}}
\newcommand{\q}[0]{Q_6}
\newcommand{\V}[1]{\boldsymbol{#1}}
\newcommand{\paren}[1]{\left( {#1} \right ) }

\newcommand{\caja}[1]{\left[ {#1} \right ] }
\newcommand{\grad}[0]{\boldsymbol{\nabla}}

\begin{document}

\title{Equilibrium fluid-solid coexistence of hard spheres}

\author{L.~A.~Fernandez} 
\author{V.~Martin-Mayor} 
\author{B.~Seoane}
  \affiliation{Departamento de F\'\i{}sica Te\'orica I, Universidad
  Complutense, 28040 Madrid, Spain.}
\affiliation{Instituto de Biocomputaci\'on y F\'{\i}sica de Sistemas
  Complejos (BIFI), Spain.}
\author{P. Verrocchio} 
\affiliation{Dipartimento di Fisica,
  Universit{\`a} di Trento, via Sommarive 14, I-38050 Povo, Trento,
  Italy.}  
\affiliation{Istituto Sistemi Complessi (ISC-CRS), UOS
  Sapienza, Via dei Taurini 19, 00185 Roma, Italy.}
\affiliation{Interdisciplinary Laboratory for Computational Phyiscs (LISC), Trento, Italy.}
\affiliation{Instituto de Biocomputaci\'on y F\'{\i}sica de Sistemas
  Complejos (BIFI), Spain.}

\begin{abstract}
We present a tethered Monte Carlo simulation of the crystallization of
hard spheres. Our method boosts the traditional umbrella sampling to the point
of making practical the study of  constrained Gibbs' free energies depending
on several crystalline order-parameters. We obtain high-accuracy estimates of
the fluid-crystal coexistence pressure for up to $2916$ particles (enough to
accommodate fluid-solid interfaces). We are able to extrapolate to infinite
volume the coexistence pressure ($p_\mathrm{co}=11.5727(10)
k_\mathrm{B}T/\sigma^3$) and the interfacial free energy
($\gamma_{\{100\}}=0.636(11) k_\mathrm{B}T/\sigma^2$).
\end{abstract}


\maketitle

Crystallization is a vast field of research, where experiments and theory
cross-fertilize. Hard-spheres (HS) provide a celebrated example: the numerical
finding of a fluid-solid phase transition~\cite{alder:57,*wood:57} motivated
experiments on colloids~\cite{pusey:86,pusey:89}. Finding an accurate
procedure to locate the equilibrium phase boundaries for HS is a crucial step
to address the self-assembly of complex
molecules~\cite{manoharan:03,*glotzer:07}, as modeled by HS plus
non-spherical interactions (e.g patchy~\cite{bianchi:06} and Janus
particles~\cite{sciortino:09}).

Up to now, numerical simulations of crystallization phase transitions have
been well behind their fluid-fluid counterpart (e.g. vapor-liquid
equilibria~\cite{allen:89}).  Actually, HS are the preferred benchmark for
numerical approaches to crystallization. Yet, the lack of exact solutions
enhances the importance of accurate numerical and/or experimental studies.

However, for preexisting numerical methods, a simulation whose starting
configuration is a fluid  never reaches the equilibrium crystal. Much as in
experiments~\cite{pusey:89}, the simulation gets stuck in a metastable
crystal, or a defective crystal (or even a glass~\cite{zaccarelli:09}). The
proliferation of metastable states defeats optimized Monte Carlo (MC) methods
that overcome free-energy barriers in simpler
systems~\cite{berg:92,wang:01,martin-mayor:07}. Besides, experimental and
numerical determinations of the interfacial free energy are plainly
inconsistent (maybe due to a small electrical charge in the colloidal
particles~\cite{anderson:02}).

Since feasible numerical methods~\cite{vega:08} could not form the correct
crystalline phase spontaneously, choosing the starting particle configuration
became an issue (e.g. crystalline or a carefully crafted mixture of solid and
fluid phases).  Methods can be classified as {\em equilibrium} or {\em
  nonequilibrium.\/} In the phase switch MC~\cite{wilding:00}, one tries to
achieve fluid-crystal equilibrium (only up to $N=500$ HS~\cite{errington:04}).
An alternative is the separate computation of the fluid and solid
free energies, supplemented with the conditions of equal pressure, temperature
and chemical potential. For the fluid's free energy, one resorts to
thermodynamic integration, while choices are available for the crystal
(Wigner-Seitz~\cite{hoover:68}, Einstein crystal~\cite{frenkel:84,polson:00},
Einstein molecule~\cite{vega:07}).  The nonequilibrium \textit{direct
  coexistence} method~\cite{ladd:77,noya:08} handles larger
systems~\cite{zykova-timan:10}.

As for the accuracy, in equilibrium computations the coexistence
pressure $\p$ was obtained with precisions of $\sim 0.1\%$ (at finite
$N$). Yet, the $N$ values that can be simulated are rather small. An
$N\to\infty$ extrapolation is mandatory, which degrades the final
accuracy to $\sim 1\%$~\cite{errington:04,wilding:00,vega:07} (results
are summarized in Table~\ref{tab:pN}). The situation
improves by an order of magnitude for the direct-coexistence method.
With the exception of~\cite{errington:04}, the different estimations
of $\p$ are compatible, although with widely differing accuracies.

The computation of the interfacial free energy, $\gamma$, is more
involved, since the issue of spatially heterogeneous mixtures of fluid
and solid can no longer be skipped (as done in equilibrium
computations of $\p$). Indeed, recent estimations are either precise
but mutually incompatible~\cite{davidchack:10,cacciuto:03}, or of
lesser accuracy~\cite{mu:05}.

Here, we introduce a tethered MC~\cite{fernandez:09,martin-mayor:11}
approach to HS crystallization. The correct crystal appears in our
simulation by constraining the value of two order parameters. At
variance with preexisting methods, the crystal found is independent
from the starting particle configuration. Tethered MC provides a major
simplification for the standard umbrella sampling
method~\cite{torrie:74,*torrie:77,*bartels:00,tenwolde:95}:
chemical-potential differences among fluid and crystal are very
precisely computed from a thermodynamic integration.  In fact, our
method resembles studies of liquid-vapor
equilibria~\cite{schrader:09,binder:11}. We go continuously from the
fluid to the crystal by varying a reaction coordinate that labels the
intermediate states.  Rather than particle density, our reaction
coordinate is a blend of bond-orientational crystal order parameters
with different
symmetries~\cite{steinhardt:83,duijneveldt:92,angioletti:10}.  Very
accurate determinations of the coexistence pressure and the
interfacial free energy follow.  The number of HS ranges $108\!\le\!
N\!\!=\!\!4 n^3\! \le \!4000$, ($n$ integer). Our largest systems do
show the surface-driven geometric transitions characteristic of the
asymptotic large $N$ regime~\cite{biskup:02,binder:03,macdowell:06}.

We consider $N$ hard spheres of diameter $\sigma$, at constant pressure $p$,
in a cubic box with periodic boundary conditions. The equilibrium crystal is
face-centered cubic (FCC)~\cite{bolhuis:97}. With the shorthand $\V{R}$ for
the particle positions, $\{\V{r}_i\}_{i=1}^N$, Gibbs free energy $g(p,T)$ is
given by
\begin{equation}\label{eq-Gibbs}
\mathrm{e}^{-N \beta g(p,T)}=\frac{p\beta}{ N!
  \Lambda^{3N}}\int_0^\infty\mathrm{d} V \mathrm{e}^{-\beta p V}\int
\mathrm{d} \V{R}\, H(\V{R})\,,
\end{equation}
($\Lambda$: de Broglie thermal wavelength, $\beta=1/(k_\mathrm{B}T)$ and
$H(\V{R})=0$ if any pair of spheres overlaps, or 1 otherwise).

We loosely constraint the values of two global order parameters, $\q$ and
$C$. The well-known $\q$ detects the spatially coherent alignment of
nearest-neighbors bonds in a lattice~\cite{steinhardt:83,duijneveldt:92}. It
is the $l=6$ instance of
\begin{equation}\label{eq:def-Q6}
 Q_l \equiv \sqrt{ \frac{4 \pi}{2l +1}\sum_{m = -l}^{l}
   \Bigg|\frac{\sum_{i=1 }^{N}\, \sum_{j=1}^{N_b(i)} Y^m_{l}({\hat
   r_{ij}})}{\sum_{i=1 }^{N} N_b(i)}\Bigg|^2 }\,,
\end{equation}
($Y^m_{l}({\hat r_{ij}})$: spherical harmonics; ${\hat r_{ij}}$: unitary
vector pointing from particle $i$ to particle $j$; $N_b(i)$: number of
neighbors of particle $i$\footnote{Particles $i$ and $j$ are
  \textit{neighbors} if $r_{ij}<1.5\ \sigma$. In the ideal FCC structure, for
  all particle-densities relevant to us, this choice includes only the nearest
  neighbors shell.}). $Q_6$ is positive in a crystal, while it is negligible
($\q\sim 1/\sqrt{N}$) in a fluid. Yet, $\q$'s rotational invariance is a
nuisance: enforcing a large $\q$ causes a crystal grain to grow in the fluid,
but its orientation in the simulation box is arbitrary.  In fact, when the
grain finally hits itself through the periodic box's boundaries, long-lived
metastable helicoidal crystals appear. The cure is an order parameter with
only cubic symmetry~\cite{angioletti:10}:
\begin{equation}
\label{eq:C}
C=\frac{2288}{79}\frac{\sum_{i=1}^{N} \sum_{j=1}^{N_b(i)}c_{\alpha}({\hat r_{ij}})}{\sum_{i=1 }^{N}N_b(i)}-\frac{64}{79} \,,
\end{equation}
where $c_{\alpha}({\hat r})=x^4y^4(1-z^4)+x^4z^4(1-y^4)+y^4z^4(1-x^4)$.  $C=1$
in an ideal, well aligned FCC, while $C\approx 0$ for a fluid.  Constraining a
large $C$ value suffices to obtain a nice crystal, irrespectively of the
starting configuration (either a gas or an FCC structure). Still, $\q$ helps
us label unambiguously the intermediate states between the fluid and the
FCC: some helicoidal crystals and the fluid-solid mixtures differ on their
$\q$ values (but not on $C$).

To enforce the quasi-constraints $C(\V{R})\approx \hc$, $\q(\V{R})\approx
\hq$~\cite{fernandez:09,martin-mayor:11}, first multiply the integrand in
Eq.~\eqref{eq-Gibbs} by
\begin{equation}
1=\frac{N\alpha}{2\pi}\int\mathrm{d}\hq \mathrm{d}\hc\,
 {\mathrm e}^{-\frac{N\alpha}{2}\bigl[\paren{\hq -
    Q_6(\V{R})}^2 +\paren{\hat{C} - C(\V{R})}^2\bigr]}\,.
\end{equation}
The tunable parameter $\alpha$ tightens the quasi-constraints (we choose
$\alpha=200$~\cite{martin-mayor:11}). Exchanging the integration order
in~\eqref{eq-Gibbs} yields
\begin{equation}\label{eq:ensemble-independence}
\mathrm{e}^{-N\beta g(p,T)}=\int
\mathrm{d}\hq\,\mathrm{d}\hc\,\mathrm{e}^{-N\varOmega_N(\hq,\hc,p)}\,,
\end{equation} 
where the \textit{effective potential}, $\varOmega_N(\hq,\hc,p)$ is given by 
\begin{equation}\label{eq:Omegadef}
\mathrm{e}^{-N\varOmega_N}=\frac{p\beta N\alpha}{2\pi\,N!
  \Lambda^{3N}}\int
\mathrm{d} \V{R}\, \mathrm{d} V\, \omega(\V{R},V;\hq,\hc,p)\,,
\end{equation}
 $\omega(\V{R},V;\hq,\hc,p)$ being the \emph{tethered}
weight~\footnote{Eq.~\eqref{eq:weight} behaves as an animal's tether: only if
  (say) $|\hq-\q(\V{R})|\gg 1/\sqrt{N\alpha}$ is the penalty large. Note as
  well that Eqs.~(\ref{eq:weight},\ref{eq:h}) generalize straightforwardly to
  the case of more than two quasi-constraints.}
\begin{equation}
\label{eq:weight}
\omega = H(\V{R})\ {\mathrm e}^{-\beta pV-\frac{N\alpha}{2}\, \caja{\paren{\hq -
    Q_6(\V{R})}^2 +\paren{\hat{C} - C(\V{R})}^2}}\,.
\end{equation}

Our method relies on Fluctuation-Dissipation
formulae~\cite{fernandez:09,martin-mayor:11}, obtained by taking derivatives
in Eq.~\eqref{eq:Omegadef}. We compute the gradient of $\varOmega_N$ at fixed
pressure from:
\begin{equation}
\label{eq:h} 
\grad \varOmega_N{(\hq,\hc)}=  \alpha   \big(\langle\hq-\q(\V{R})\rangle\,,\,\langle\hc-C(\V{R})\rangle\big)\,.
\end{equation}
Coordinates $(\hq^*,\hc^*)$ of local minima of $\varOmega$ are located through
$\grad \varOmega_N=0$. Furthermore, differences
$\varOmega_N({\hq}^{b},\hc^{b})-\varOmega_N(\hq^{a},\hc^{a})$ at fixed $p$ are
computed as the line integral of $\grad \varOmega_N$ along any convenient path
joining $({\hq}^{a},\hc^{a})$ with $(\hq^{b},\hc^{b})$ in the $(\hq,\hc)$
plane.

The chemical potential $g(p,T)$ is obtained from a saddle-point expansion in
Eq.~\eqref{eq:ensemble-independence}. Up to corrections vanishing as $1/N$,
$\beta g(p,T)$ is the absolute minimum of $\varOmega_N(p,\hq,\hc)$.  Yet,
close to phase coexistence, $\varOmega_N$ has two relevant minima (i.e. the
fluid and the FCC crystal). Therefore, the coexistence pressure
$p_\mathrm{co}^{(N)}$ follows from
$\varOmega_N^\mathrm{fluid}=\varOmega_N^\mathrm{FCC}$ (i.e. equal chemical
potential).

Our Metropolis MC simulation follows standard methods~\cite{allen:89}.  We
recast $\omega$ in Eq.~\eqref{eq:weight} as the Boltzmann factor for HS at
fixed pressure with a {\em fictive} potential energy $k_\mathrm{B} T N\alpha\,
[(\hq - Q_6(\V{R}))^2 +(\hat{C} - C(\V{R}))^2]/2$. Since $Q_6(\V{R})$ and
$C(\V{R})$ are built out of sums of local terms, the number of operations
needed to compute their changes after a single-particle displacement does not
grow with $N$.

\begin{figure}
\includegraphics[angle=270,width=\columnwidth,trim=23 30 30
  20]{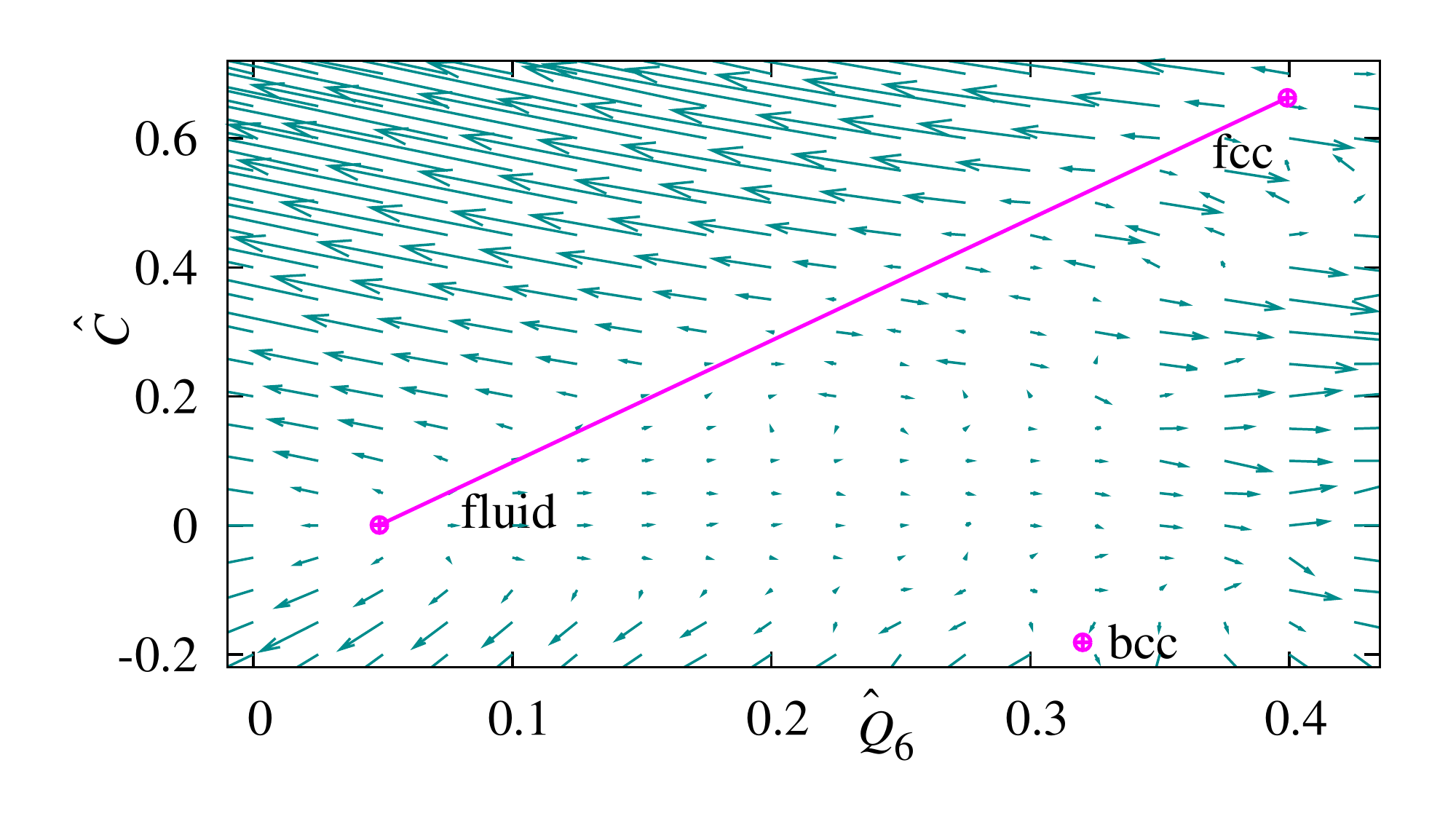}
\caption{(Color online) Vector field $\grad \varOmega_N$ as computed from
  Eq.~(\ref{eq:h}), for a system of $N=256$ hard-spheres, at the coexistence
  pressure for the fluid-FCC phase transition (we scaled $\grad \varOmega_N$
  with a factor $1/\alpha$). Both the fluid and the FCC crystal are local
  minima of the effective potential, where $\grad \varOmega_N=0$. The BCC
  coordinates are from $N=250$.}
\label{fig:grid}
\end{figure}

Our framework is illustrated in Fig.~\ref{fig:grid}, where we show $\grad
\varOmega_N(\hq,\hc)$ at $p=\p^{(N)}$.  We identify two local minima where
$\grad \varOmega_N=0$ [the fluid, close to $(\hq,\hc)=(1/\sqrt{N},0)$, and the
FCC minimum where both parameters are positive].  Note their distance to other
local minima of $\varOmega_N$, such as the body centered cubic (BCC).

\begin{figure}
\includegraphics[angle=270,width=\columnwidth,trim= 20 20 20  20]{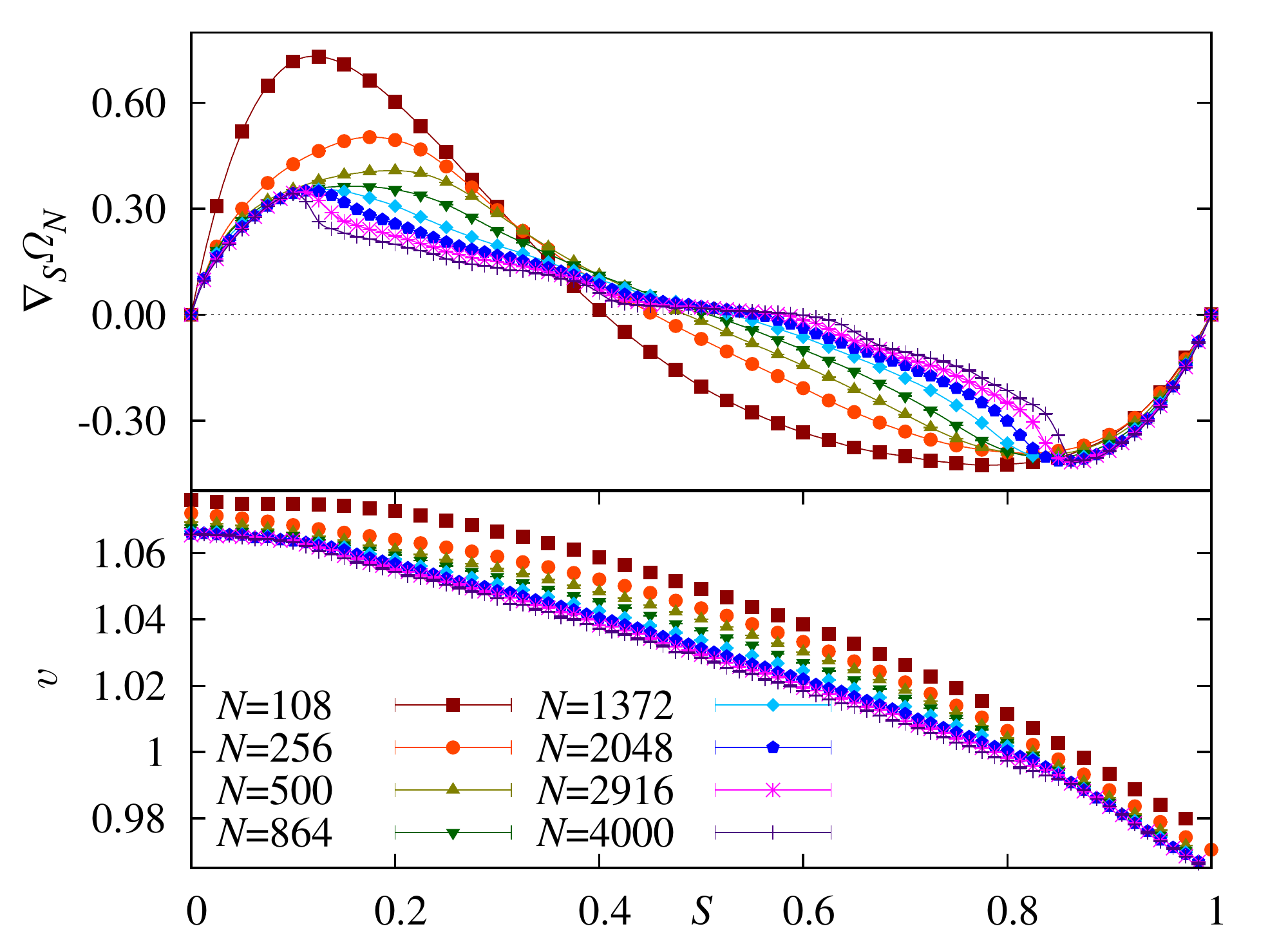} 
\caption{(Color online) ({\bf Top}) $\grad \varOmega_N$ projected over the
  liquid-FCC line, $\grad_S \varOmega_N$, vs. the line parameter $S$ ($S=0$:
  fluid, $S=1$: FCC), for all our system sizes at the simulation
  pressures. ({\bf Bottom}) Specific volume $v=V/N$ as a function of line
  parameter $S$. At large $N$, $v$ becomes a linear function, as expected for
a convex combination of pure phases~\cite{ruelle:69}.}
\label{fig:h}
\end{figure}

Our main goal is to compute
$\Delta\varOmega(p)\!=\!\varOmega^\mathrm{FCC}-\varOmega^\mathrm{fluid}$,
choosing the straight segment in Fig.~\ref{fig:grid} as integration
path. The path is parameterized by our \emph{reaction coordinate}, $S$
($S\!=\!0$: fluid, $S\!=\!1$: FCC). Actually, due to the {\em
  additivity} of $\q$ and $C$, choosing this segment is a must if we
are to compute the interfacial free energy \footnote{A magnitude $A$
  is {\em additive} if $N A$ is extensive: gluing together systems 1,2
  (with $N^{(i)}$ particles and $A=A^{(i)}$, $i=1,2$), results in a
  total system with $N=N^{(1)}+N^{(2)}$ particles and $NA=N^{(1)}
  A^{(1)} +N^{(2)}A^{(2)}$ (plus subdominant corrections such as
  surface effects $\sim N^{2/3}$).  $C$ is additive to a great
  accuracy for coexisting fluid and FCC phases, because the average
  number of neighbors $N_b$ is very similar in both phases ($5\%$
  difference, with negligible effects on additivity in our $N$ range,
  as compared with surface effects). $\q$ is additive
  only if one of the subsystems, say $i=1$, is a liquid so that
  $\q^{(1)}\sim 1/\sqrt{N^{(1)}}$ ($\q$ is a pseudo-order parameter,
  i.e. a strictly positive quantity which is of order $1/\sqrt{N}$ in
  a disordered phase). For studies of interfaces on larger systems, it
  would be advisable to choose exactly additive order parameters.}.
Indeed, physical fluid-solid coexistence is a convex combination of
the two pure phases~\cite{ruelle:69}, which provides a physical
interpretation for $S$ as the fraction of particles in the coexisting
solid phase: in the large $N$ limit, $v$, $C$ and $\q$ vary linearly
with $S$ (see Fig.~\ref{fig:h}---bottom).

Our simulation set up is as follows. We start by locating $(\hq,\hc)$ for the
FCC and liquid minima at $p\approx \p^{(N)}$. The first guess is obtained from
$NpT$ simulations with crystalline/disordered starting configurations. We
later refine by solving for $\grad\varOmega_N=0$~\cite{martin-mayor:11}.

Next, we introduce a uniform $S$ grid on the liquid-FCC line and
perform {\em independent} simulations at fixed $(\hq,\hc,p)$ (see
Appendix \ref{app:extended} for simulation details). As a test for
equilibration, achieved for all $N$ but $N=4000$, every run was
performed twice (starting from an ideal gas or from an ideal FCC
crystal)~\footnote{Our runs for $N\leq 2916$ are, at least, $100 \tau$
  long ($\tau$ is the integrated autocorrelation time~\cite{sokal:97},
  computed for $\q$ and $v$~\cite{martin-mayor:11}). For $N=2916$, but
  only at $S=0.4$, we find metastability with a helicoidal
  configuration (however, its contribution to final quantities is
  smaller than statistical errors).  Metastabilities arise often for
  $N=4000$, at intermediate $S$ (yet, a careful selection of starting
  configurations yields a $\grad\varOmega_N$ with smooth $S$
  dependency).}.

\begin{figure}
\includegraphics[angle=270,width=\columnwidth,trim= 20 20 20  20]{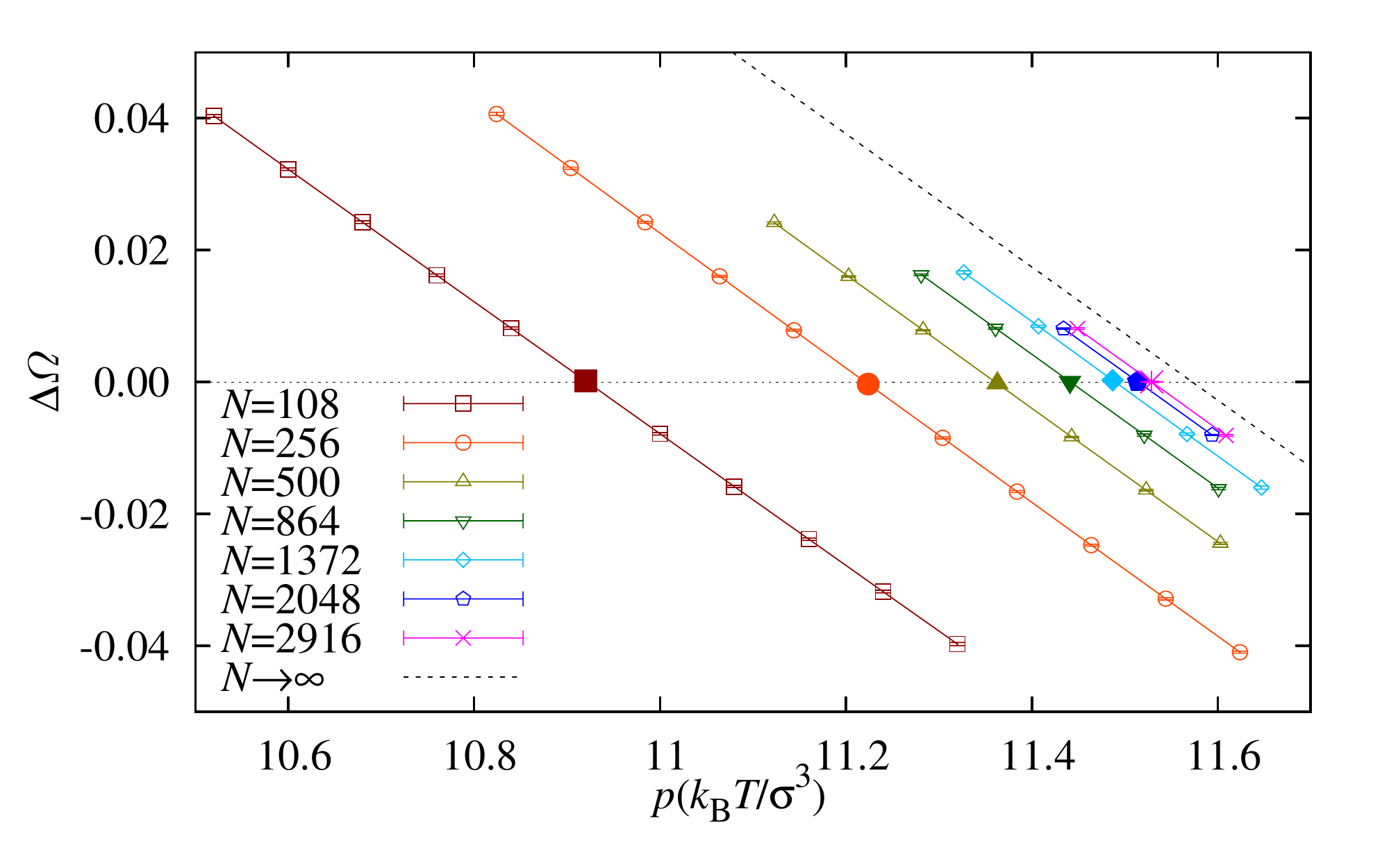} 
\caption{(Color online) Effective-potential difference
  $\Delta\varOmega(p)\!=\!\varOmega^\mathrm{FCC}-\varOmega^\mathrm{fluid}$,
  as a function of pressure. At $p_\mathrm{co}^{(N)}$, $\Delta
  \varOmega_N=0$. The large $N$ limit stems from
  $\Delta\varOmega(p)=(v^\mathrm{FCC}-v^\mathrm{fluid})(p-p_\mathrm{co})/(k_\mathrm{B}T)
  + {\cal O}((p-p_\mathrm{co})^2)$. The simulated
  pressures (see Table~\ref{tab:pN}) correspond to the larger, filled
  symbols.}
\label{fig:omega}
\end{figure}

Now, at variance with umbrella sampling, $\Delta\varOmega(p)$ follows from the
integral over $0\leq S\leq 1$ of $\grad_S\varOmega_N$, the projection of
$\grad \varOmega_N$ along the straight-line, Fig.~\ref{fig:h}---top.  We use
\textit{reweighting} extrapolations~\cite{ferrenberg:88,martin-mayor:11} to
obtain $\Delta\varOmega(p)$ as a function of pressure. Then, it is easy to
locate $\p^{(N)}$, Fig.~\ref{fig:omega}. Statistical errors are estimated as
in~\cite{martin-mayor:07}.

We obtain $\p\!=\!11.5727(10)$ in units of $k_\mathrm{B}T/\sigma^2$,
in the large-$N$ limit. This result is six times more accurate than
the best nonequilibrium estimate,
$\p\!=\!11.576(6)$~\cite{zykova-timan:10} and improves by a factor of
90 over the equilibrium estimate, $\p\!=\!11.49(9)$~\cite{wilding:00}.
We compute $\p$ through a fair fit ($\chi^2=2.61$ for three degrees of
freedom) of the $\p^{(N)}$ listed in Table~\ref{tab:pN} to a second
order polynomial in $1/N$~\cite{borgs:92}.

As for the interfacial free energy, $\gamma$, we need to consider
inhomogeneous configurations~\footnote{The tethering approach should not
  induce artificial interfaces. In fact, mathematically, the
  interfacial free-energy is defined though the ratio of two partition
  functions with different boundary conditions. The tethered potential
  does not change the partition function [with any boundary
    conditions, see Eq.~\eqref{eq:ensemble-independence}].}. In fact,
due to the periodic boundary conditions, at intermediate $S$ the
surface energy is minimized by mixed configurations where a
crystalline slab (or cylinder, or bubble) is surrounded by fluid, see
the snapshots in Appendix~\ref{app:geometric}.  As in vapor-liquid
equilibria~\cite{macdowell:06,binder:11}, transitions among different
geometries arise when $S$ is varied.  These transitions result in the
cusps and steps that appear for large $N$ in $\grad_S\varOmega_N$,
Fig.~\ref{fig:h}---top, and can be detected as well through the
fluctuations of the particle density~\cite{martin-mayor:11}. Under
these circumstances, $\gamma$ may be computed using Binder's
method~\cite{binder:82}. The effective potential has a local maximum
along the line that joins the FCC and the fluid (the solution of
$\grad_S\varOmega_N=0$ at $S^*\approx 0.5$,
Fig.~\ref{fig:h}---top). The excess free energy is due to the {\em
  two} interfaces that the fluid presents with a crystalline slab
parallel to the simulation box ($\{100\}$ planes). Then the
interfacial free energy at $\p^{(N)}$ is
\begin{equation}
\gamma^{(N)}=
k_\mathrm{B}T\,N\paren{\varOmega_{s^*}-\varOmega_\text{FCC}}/(2
\mean{Nv}_{S^*}^{2/3})\,.
\end{equation}
The $\gamma^{(N)}$ (listed in Table~\ref{tab:pN}) are extrapolated
as~\cite{billoire:94}
\begin{equation}\label{eq:extrapol-gamma}
\frac{\gamma^{(N)}\sigma^2}{k_\mathrm{B}T}=\frac{\gamma\sigma^2}
      {k_\mathrm{B}T}+ \frac{a_2-\mathrm{log} N}{6N^{2/3}}+
\frac{a_3}{N}+\frac{a_4}{N^{4/3}}+\ldots
\end{equation}
A fit for $256\le N\le 2916$ yields $\gamma=0.636(11)$ in units of
$k_\mathrm{B}T/\sigma^2$ ($\chi^2=0.14$ for two degrees of freedom). We remark
that the difference among the fit and $\gamma^{(N=4000)}$ is one fifth of the
error bar (Table~\ref{tab:pN}). Also, the extrapolation for $500\le N\le 2916$ merely doubles the
final error estimate. Our result is compatible with
$\gamma=0.64(2)$~\cite{mu:05} and $\gamma=0.619(3)$~\cite{cacciuto:03}, but
not with $\gamma=0.5820(19)$~\cite{davidchack:10}. We remark that the
$\gamma^{(N)}$ estimation is fairly sensitive to $p$~\cite{martin-mayor:11},
an effect not systematically considered
in~\cite{davidchack:10,cacciuto:03,mu:05}. Note that
Eq.~\eqref{eq:extrapol-gamma} holds if $\gamma^{(N)}$ is computed at
$\p^{(N)}$.

A final warning is in order. Not much is known about the effect of the
$\grad_S \varOmega_N$'s cusps and steps, Fig.~\ref{fig:h}---top, in
the large-$N$ extrapolation $\gamma^{(N)}\to\gamma$.  This
non-smoothness is a consequence of the geometric transitions that
arise in our larger systems. However, the analogy with simpler
models~\cite{martin-mayor:07} (e.g. the $D\!=\!2$ Potts model, where
comparison with exact solutions is possible), strongly suggests that
these cusps and steps are inconsequential for the $\p^{(N)}\to\p$
extrapolation.

\begin{table*}[t]
\centering
\begin{tabular*}{\textwidth}{@{\extracolsep{\fill}}cccllllllllcll}
\cline{1-14} 
& & &\multicolumn{4}{c}{This work} && \multicolumn{1}{c}{\cite{wilding:00}}&\multicolumn{1}{c}{\cite{errington:04}}&&\cite{zykova-timan:10}&&\multicolumn{1}{c}{\cite{vega:07}}\\
 $N$ & $p^\mathrm{simulation}$&
&$\mean{v}^\text{FCC}$&$\mean{v}^\text{fluid}$&$\gamma_{\{100\}}$&$\p$&&\multicolumn{2}{c}{Phase switch}&
&Direct coexistence&&\multicolumn{1}{c}{E. M.}\\
\cline{1-2}\cline{4-7}\cline{9-10}\cline{12-12}\cline{14-14}
108 & 10.92 & &0.97580(7) &1.07611(8) &0.4063(12)&10.9216(18)& &10.94(4)
&11.00(6)& & & &11.02(5)     \\ 
256 & 11.224& &0.97049(6) &1.07202(7) &0.4243(8) &11.2209(13)& &11.23(4)&11.25(1)&&&&11.26(5) \\ 
500 & 11.363& &0.96796(10)& 1.06932(7)&0.4798(8) &11.3607(8) & &        &11.34(1)&&&&11.35(3) \\ 
864 & 11.441& &0.96796(10)& 1.06932(7)&0.5285(12)&11.4416(13)& &        &        &         &&&     \\
1372& 11.487& &0.96549(14)&1.06659(13)&0.5611(14)&11.4897(13)& &        &        & &&&11.50(3)\\ 
2048& 11.514& &0.96500(14)&1.06577(15)&0.5832(10)&11.5146(7) & &        &        & &&&11.52(3)\\ 
2916& 11.529& &0.96468(14)&1.06545(19)&0.5971(12)&11.5311(15)& &        &        &         &&&     \\
\cline{1-2}\cline{4-7}
4000& 11.54 & &0.96461(13)&1.06556(15)&0.607(2)&11.5452(11)\\
\cline{1-2}\cline{4-7}\cline{9-10}\cline{12-12}\cline{14-14} 
$\infty$&& &0.96405(3)&1.06448(10)&0.636(11)&11.5727(10)&&11.49(9)&11.43(2)&&11.576(6)&&11.54(4)\\ 
\multicolumn{2}{c}{$\chi^2/$degrees of freedom}& &$0.32/3$&$0.61/2$&0$.14/2$&$2.61/3$\\
\cline{1-14}
\end{tabular*}
\caption{For each $N$, we report the simulated pressure in units of
  $k_\mathrm{B}T/\sigma^3$, the specific volume of the coexisting
  phases, the \{100\} surface tension $\gamma_{\{100\}}$ (in
  $k_\mathrm{B}T/\sigma^2$ units) and the phase-coexistence pressure
  $p_\mathrm{co}^N$ (which is compared with work by other authors
  using different methods: phase switch Monte Carlo, the
  non-equilibrium direct coexistence method, and the Einstein Molecule
  approach).  We extrapolate $p_\mathrm{co}^{(N)}$ to the large $N$
  limit as $p_\mathrm{co}^{(N)}= p_\mathrm{co}^{\infty} +a_1/N +
  a_2/N^2$ for $256\! \leq\! N\!  \leq\! 2916$
  ($p_\mathrm{co}^{N=4000}$ is compatible but not included in the fit
  because of dubious equilibration). The specific volume was
  extrapolated linearly in $1/N$ ($N\!\geq\! 256$ for the FCC and
  $N\geq 500$ for the fluid).}
\label{tab:pN}
\end{table*} 

In summary, we have introduced a tethered
MC~\cite{fernandez:09,martin-mayor:11} approach to HS crystallization.
We go continuously from the fluid to the crystal by varying a reaction
coordinate.  Tethered MC provides a major simplification to umbrella
sampling, which makes it possible to study multi-constrained free
energies.  At variance with previous methods, our simulations
equilibrate (i.e.  we find results independent of the starting
particle configuration), not only for the formation of the
space-filling crystal, but even for the more difficult case of mixed
states with fluid-crystal interfaces.  Our estimation of the
coexistence pressure is, by far, the most accurate to date. That of
the interfacial free energy is compatible with most (but not all)
recent determinations. Should one wish to reach larger $N$, the
tethered strategy would easily accommodate additional order
parameters.  The method can also be generalized to other simple
liquids, or to investigate the glass transition.

\begin{acknowledgments}
We thank K. Binder, C. de Vega, L.G. MacDowell, B. Lucini and D. Yllanes for
enlightening discussions. Simulations were carried out at BIFI.  We
acknowledge support from MICINN, Spain, through research contracts
FIS2009-12648-C03, FIS2008-01323 and from UCM-Banco de Santander. B.S. was
supported by FPU program.
\end{acknowledgments}

\appendix

\section{Extended simulation details}\label{app:extended}

We provide some additional details for the interested reader. In
particular, we give information necessary to reproduce our analysis
and/or our simulations.

As shown in the Fig. 2, we take a segment of the
straight line in the $(\hq,\hc)$ plane that joins the fluid and the
FCC minima of the effective potential. This segment is divided evenly
in a grid of $N_S$ points ($N_S\!=\!42$ for $N\!\leq\! 1372$, while
$N_S\!=\!82$ for $N\!\geq\!  2048$). All $N_S$ points are simulated at
the same pressure $p$ (see Table~\ref{tab:pN}). We selected these $p$
values by means of short, preliminary simulations.

At each of the $N_S$ $(\hq,\hc,p)$ points, we run two independent
simulations with different initial conditions, an ideal gas and a
perfect FCC crystal. Each of the $2\times N_S$ runs had a length of
$10^6$ MC steps (1 MC step is composed of $N$ particle displacements
followed by one volume-change attempt). In addition, we show in
Table~\ref{tab:pN} 
some $N-$dependent observables computed in the
simulation, as well as the large-$N$ extrapolation. In order to ease
comparison, we also tabulate the $\p^{(N)}$ values obtained by other
groups (using different approaches).

\section{Geometrical transitions}\label{app:geometric}
\begin{figure*}
\includegraphics[angle=270,width=\columnwidth,trim=20 100 10 100]{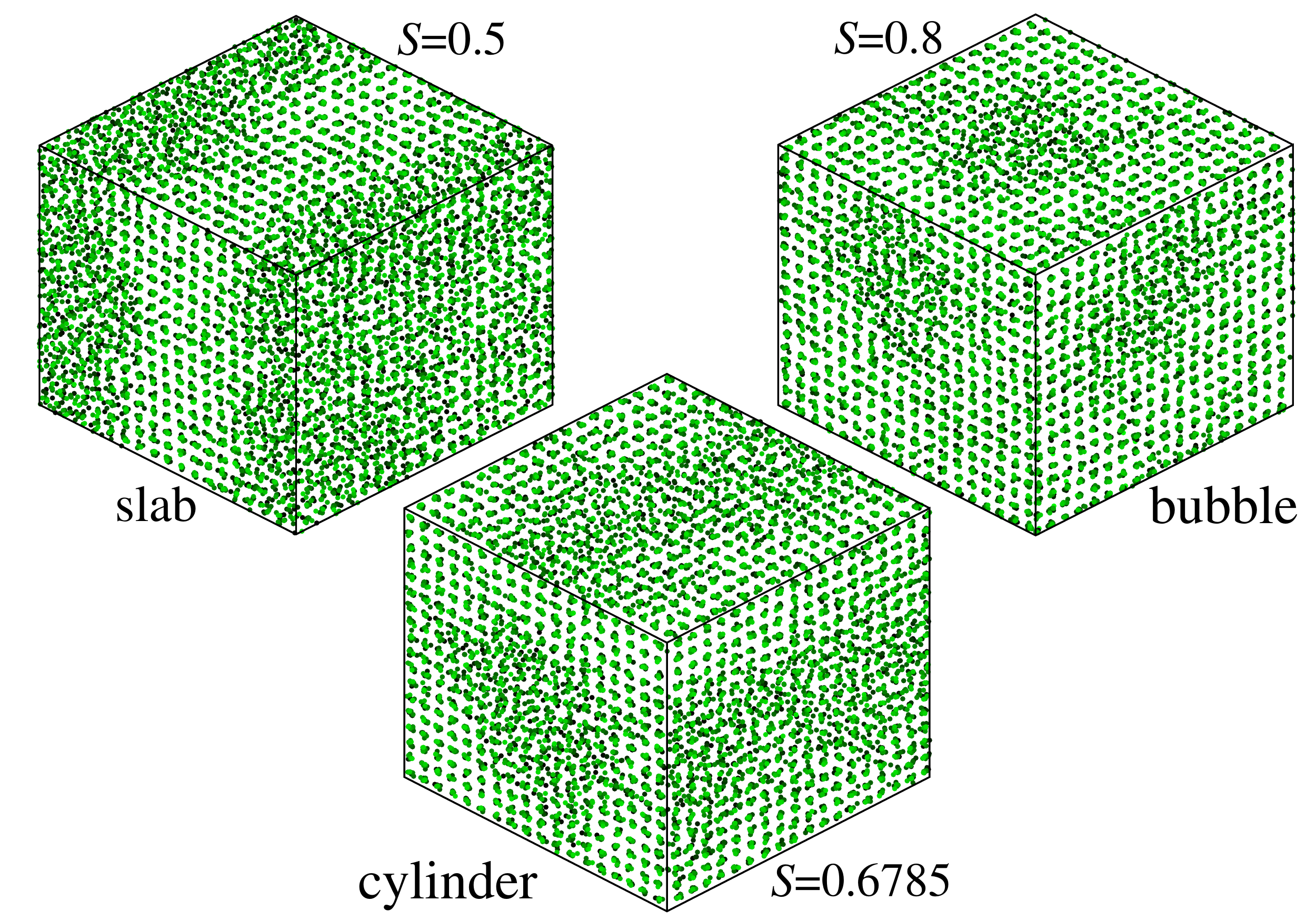}
\caption{Snapshots of mixed configurations for $N=2916$ particles
  found as the line parameter $S$ varies. We present projections in
  the three Cartesian directions. To improve visibility, the radii are
  a fraction of the real ones, and the darkness is an increasing function of
  the distance to the projection plane.}
\label{fig:configuraciones}
\end{figure*}

As discussed in the main text, in a system with periodic boundary
conditions, geometrical transitions arise when the line parameter $S$
varies from the liquid to the solid. In fact, the system struggles to
minimize the surface energy while respecting the global constraints
for $\q$ and $C$. Depending on the fraction of crystal phase, which is
fixed by $S$, the minimizing geometry can be either a bubble, a
cylinder or a slab of liquid in a crystal matrix (or vice versa). An
example of each type of configuration is displayed in
Fig. \ref{fig:configuraciones}. As $S$ varies, the minimizing geometry
changes at definite $S$ values. This phenomenon is named {\em
  geometric transition}, and has been previously studied in simpler
models (for instance, first-order transitions in lattice magnetic
systems, or fluid-gas phase-coexistence).

\begin{thebibliography}{54}%
\makeatletter
\providecommand \@ifxundefined [1]{%
 \@ifx{#1\undefined}
}%
\providecommand \@ifnum [1]{%
 \ifnum #1\expandafter \@firstoftwo
 \else \expandafter \@secondoftwo
 \fi
}%
\providecommand \@ifx [1]{%
 \ifx #1\expandafter \@firstoftwo
 \else \expandafter \@secondoftwo
 \fi
}%
\providecommand \natexlab [1]{#1}%
\providecommand \enquote  [1]{``#1''}%
\providecommand \bibnamefont  [1]{#1}%
\providecommand \bibfnamefont [1]{#1}%
\providecommand \citenamefont [1]{#1}%
\providecommand \href@noop [0]{\@secondoftwo}%
\providecommand \href [0]{\begingroup \@sanitize@url \@href}%
\providecommand \@href[1]{\@@startlink{#1}\@@href}%
\providecommand \@@href[1]{\endgroup#1\@@endlink}%
\providecommand \@sanitize@url [0]{\catcode `\\12\catcode `\$12\catcode
  `\&12\catcode `\#12\catcode `\^12\catcode `\_12\catcode `\%12\relax}%
\providecommand \@@startlink[1]{}%
\providecommand \@@endlink[0]{}%
\providecommand \url  [0]{\begingroup\@sanitize@url \@url }%
\providecommand \@url [1]{\endgroup\@href {#1}{\urlprefix }}%
\providecommand \urlprefix  [0]{URL }%
\providecommand \Eprint [0]{\href }%
\providecommand \doibase [0]{http://dx.doi.org/}%
\providecommand \selectlanguage [0]{\@gobble}%
\providecommand \bibinfo  [0]{\@secondoftwo}%
\providecommand \bibfield  [0]{\@secondoftwo}%
\providecommand \translation [1]{[#1]}%
\providecommand \BibitemOpen [0]{}%
\providecommand \bibitemStop [0]{}%
\providecommand \bibitemNoStop [0]{.\EOS\space}%
\providecommand \EOS [0]{\spacefactor3000\relax}%
\providecommand \BibitemShut  [1]{\csname bibitem#1\endcsname}%
\let\auto@bib@innerbib\@empty
\bibitem [{\citenamefont {Alder}\ and\ \citenamefont
  {Wainwright}(1957)}]{alder:57}%
  \BibitemOpen
  \bibfield  {author} {\bibinfo {author} {\bibfnamefont {B.~J.}\ \bibnamefont
  {Alder}}\ and\ \bibinfo {author} {\bibfnamefont {T.~E.}\ \bibnamefont
  {Wainwright}},\ }\href@noop {} {\bibfield  {journal} {\bibinfo  {journal} {J.
  Chem. Phys.}\ }\textbf {\bibinfo {volume} {27}},\ \bibinfo {pages} {1208}
  (\bibinfo {year} {1957})}\BibitemShut {NoStop}%
\bibitem [{\citenamefont {Wood}\ and\ \citenamefont
  {Jacobson}(1957)}]{wood:57}%
  \BibitemOpen
  \bibfield  {author} {\bibinfo {author} {\bibfnamefont {W.~W.}\ \bibnamefont
  {Wood}}\ and\ \bibinfo {author} {\bibfnamefont {J.~D.}\ \bibnamefont
  {Jacobson}},\ }\href@noop {} {\bibfield  {journal} {\bibinfo  {journal} {J.
  Chem. Phys.}\ }\textbf {\bibinfo {volume} {27}},\ \bibinfo {pages} {1207}
  (\bibinfo {year} {1957})}\BibitemShut {NoStop}%
\bibitem [{\citenamefont {Pusey}\ and\ \citenamefont {van
  Megen}(1986)}]{pusey:86}%
  \BibitemOpen
  \bibfield  {author} {\bibinfo {author} {\bibfnamefont {P.~N.}\ \bibnamefont
  {Pusey}}\ and\ \bibinfo {author} {\bibnamefont {van Megen}},\ }\href@noop {}
  {\bibfield  {journal} {\bibinfo  {journal} {Nature}\ }\textbf {\bibinfo
  {volume} {320}},\ \bibinfo {pages} {340} (\bibinfo {year}
  {1986})}\BibitemShut {NoStop}%
\bibitem [{\citenamefont {Pusey}\ \emph {et~al.}(1989)\citenamefont {Pusey},
  \citenamefont {van Megen}, \citenamefont {Bartlett}, \citenamefont
  {Ackerson}, \citenamefont {Rarity},\ and\ \citenamefont
  {Underwood}}]{pusey:89}%
  \BibitemOpen
  \bibfield  {author} {\bibinfo {author} {\bibfnamefont {P.~N.}\ \bibnamefont
  {Pusey}}, \bibinfo {author} {\bibfnamefont {W.}~\bibnamefont {van Megen}},
  \bibinfo {author} {\bibfnamefont {P.}~\bibnamefont {Bartlett}}, \bibinfo
  {author} {\bibfnamefont {B.~J.}\ \bibnamefont {Ackerson}}, \bibinfo {author}
  {\bibfnamefont {J.~G.}\ \bibnamefont {Rarity}}, \ and\ \bibinfo {author}
  {\bibfnamefont {S.~M.}\ \bibnamefont {Underwood}},\ }\href@noop {} {\bibfield
   {journal} {\bibinfo  {journal} {Phys. Rev. Lett.}\ }\textbf {\bibinfo
  {volume} {63}},\ \bibinfo {pages} {2753} (\bibinfo {year}
  {1989})}\BibitemShut {NoStop}%
\bibitem [{\citenamefont {Manoharan}\ \emph {et~al.}(2003)\citenamefont
  {Manoharan}, \citenamefont {Elsesser},\ and\ \citenamefont
  {Pine}}]{manoharan:03}%
  \BibitemOpen
  \bibfield  {author} {\bibinfo {author} {\bibfnamefont {A.~B.}\ \bibnamefont
  {Manoharan}}, \bibinfo {author} {\bibfnamefont {M.~T.}\ \bibnamefont
  {Elsesser}}, \ and\ \bibinfo {author} {\bibfnamefont {D.~J.}\ \bibnamefont
  {Pine}},\ }\href@noop {} {\bibfield  {journal} {\bibinfo  {journal}
  {Science}\ }\textbf {\bibinfo {volume} {301}},\ \bibinfo {pages} {483}
  (\bibinfo {year} {2003})}\BibitemShut {NoStop}%
\bibitem [{\citenamefont {Glotzer}\ and\ \citenamefont
  {Solomon}(2007)}]{glotzer:07}%
  \BibitemOpen
  \bibfield  {author} {\bibinfo {author} {\bibfnamefont {S.}~\bibnamefont
  {Glotzer}}\ and\ \bibinfo {author} {\bibfnamefont {M.~J.}\ \bibnamefont
  {Solomon}},\ }\href@noop {} {\bibfield  {journal} {\bibinfo  {journal} {Nat.
  Mater.}\ }\textbf {\bibinfo {volume} {6}},\ \bibinfo {pages} {557} (\bibinfo
  {year} {2007})}\BibitemShut {NoStop}%
\bibitem [{\citenamefont {Bianchi}\ \emph {et~al.}(2006)\citenamefont
  {Bianchi}, \citenamefont {Largo}, \citenamefont {Tartaglia}, \citenamefont
  {Zaccarelli},\ and\ \citenamefont {Sciortino}}]{bianchi:06}%
  \BibitemOpen
  \bibfield  {author} {\bibinfo {author} {\bibfnamefont {E.}~\bibnamefont
  {Bianchi}}, \bibinfo {author} {\bibfnamefont {J.}~\bibnamefont {Largo}},
  \bibinfo {author} {\bibfnamefont {P.}~\bibnamefont {Tartaglia}}, \bibinfo
  {author} {\bibfnamefont {E.}~\bibnamefont {Zaccarelli}}, \ and\ \bibinfo
  {author} {\bibfnamefont {F.}~\bibnamefont {Sciortino}},\ }\href {\doibase
  10.1103/PhysRevLett.97.168301} {\bibfield  {journal} {\bibinfo  {journal}
  {Phys. Rev. Lett.}\ }\textbf {\bibinfo {volume} {97}},\ \bibinfo {pages}
  {168301} (\bibinfo {year} {2006})}\BibitemShut {NoStop}%
\bibitem [{\citenamefont {Sciortino}\ \emph {et~al.}(2009)\citenamefont
  {Sciortino}, \citenamefont {Giacometti},\ and\ \citenamefont
  {Pastore}}]{sciortino:09}%
  \BibitemOpen
  \bibfield  {author} {\bibinfo {author} {\bibfnamefont {F.}~\bibnamefont
  {Sciortino}}, \bibinfo {author} {\bibfnamefont {A.}~\bibnamefont
  {Giacometti}}, \ and\ \bibinfo {author} {\bibfnamefont {G.}~\bibnamefont
  {Pastore}},\ }\href@noop {} {\bibfield  {journal} {\bibinfo  {journal} {Phys.
  Rev. Lett.}\ }\textbf {\bibinfo {volume} {103}},\ \bibinfo {pages} {237801}
  (\bibinfo {year} {2009})}\BibitemShut {NoStop}%
\bibitem [{\citenamefont {Allen}\ and\ \citenamefont
  {Tildesley}(1989)}]{allen:89}%
  \BibitemOpen
  \bibfield  {author} {\bibinfo {author} {\bibfnamefont {M.~P.}\ \bibnamefont
  {Allen}}\ and\ \bibinfo {author} {\bibfnamefont {D.~J.}\ \bibnamefont
  {Tildesley}},\ }\href@noop {} {\emph {\bibinfo {title} {Computer Simulation
  of Liquids}}},\ \bibinfo {edition} {2nd}\ ed.\ (\bibinfo  {publisher} {Oxford
  University Press},\ \bibinfo {address} {New York},\ \bibinfo {year}
  {1989})\BibitemShut {NoStop}%
\bibitem [{\citenamefont {Zaccarelli}\ \emph {et~al.}(2009)\citenamefont
  {Zaccarelli}, \citenamefont {Valeriani}, \citenamefont {Sanz}, \citenamefont
  {Poon}, \citenamefont {Cates},\ and\ \citenamefont {Pusey}}]{zaccarelli:09}%
  \BibitemOpen
  \bibfield  {author} {\bibinfo {author} {\bibfnamefont {E.}~\bibnamefont
  {Zaccarelli}}, \bibinfo {author} {\bibfnamefont {C.}~\bibnamefont
  {Valeriani}}, \bibinfo {author} {\bibfnamefont {E.}~\bibnamefont {Sanz}},
  \bibinfo {author} {\bibfnamefont {W.~C.~K.}\ \bibnamefont {Poon}}, \bibinfo
  {author} {\bibfnamefont {M.~E.}\ \bibnamefont {Cates}}, \ and\ \bibinfo
  {author} {\bibfnamefont {P.~N.}\ \bibnamefont {Pusey}},\ }\href@noop {}
  {\bibfield  {journal} {\bibinfo  {journal} {Phys. Rev. Lett.}\ }\textbf
  {\bibinfo {volume} {103}},\ \bibinfo {eid} {135704} (\bibinfo {year}
  {2009})}\BibitemShut {NoStop}%
\bibitem [{\citenamefont {Berg}\ and\ \citenamefont {Neuhaus}(1992)}]{berg:92}%
  \BibitemOpen
  \bibfield  {author} {\bibinfo {author} {\bibfnamefont {B.~A.}\ \bibnamefont
  {Berg}}\ and\ \bibinfo {author} {\bibfnamefont {T.}~\bibnamefont {Neuhaus}},\
  }\href@noop {} {\bibfield  {journal} {\bibinfo  {journal} {Phys. Rev. Lett.}\
  }\textbf {\bibinfo {volume} {68}},\ \bibinfo {pages} {9} (\bibinfo {year}
  {1992})}\BibitemShut {NoStop}%
\bibitem [{\citenamefont {Wang}\ and\ \citenamefont {Landau}(2001)}]{wang:01}%
  \BibitemOpen
  \bibfield  {author} {\bibinfo {author} {\bibfnamefont {F.}~\bibnamefont
  {Wang}}\ and\ \bibinfo {author} {\bibfnamefont {D.~P.}\ \bibnamefont
  {Landau}},\ }\href@noop {} {\bibfield  {journal} {\bibinfo  {journal} {Phys.
  Rev. Lett.}\ }\textbf {\bibinfo {volume} {86}},\ \bibinfo {pages} {2050}
  (\bibinfo {year} {2001})}\BibitemShut {NoStop}%
\bibitem [{\citenamefont {Martin-Mayor}(2007)}]{martin-mayor:07}%
  \BibitemOpen
  \bibfield  {author} {\bibinfo {author} {\bibfnamefont {V.}~\bibnamefont
  {Martin-Mayor}},\ }\href@noop {} {\bibfield  {journal} {\bibinfo  {journal}
  {Phys. Rev. Lett.}\ }\textbf {\bibinfo {volume} {98}},\ \bibinfo {pages}
  {137207} (\bibinfo {year} {2007})}\BibitemShut {NoStop}%
\bibitem [{\citenamefont {Anderson}\ and\ \citenamefont
  {Lekkerkerker}(2002)}]{anderson:02}%
  \BibitemOpen
  \bibfield  {author} {\bibinfo {author} {\bibfnamefont {V.~J.}\ \bibnamefont
  {Anderson}}\ and\ \bibinfo {author} {\bibfnamefont {H.~N.~W.}\ \bibnamefont
  {Lekkerkerker}},\ }\href@noop {} {\bibfield  {journal} {\bibinfo  {journal}
  {Nature}\ }\textbf {\bibinfo {volume} {416}},\ \bibinfo {pages} {811}
  (\bibinfo {year} {2002})}\BibitemShut {NoStop}%
\bibitem [{\citenamefont {Vega}\ \emph {et~al.}(2008)\citenamefont {Vega},
  \citenamefont {Sanz}, \citenamefont {Abascal},\ and\ \citenamefont
  {Noya}}]{vega:08}%
  \BibitemOpen
  \bibfield  {author} {\bibinfo {author} {\bibfnamefont {C.}~\bibnamefont
  {Vega}}, \bibinfo {author} {\bibfnamefont {E.}~\bibnamefont {Sanz}}, \bibinfo
  {author} {\bibfnamefont {J.~L.~F.}\ \bibnamefont {Abascal}}, \ and\ \bibinfo
  {author} {\bibfnamefont {E.~G.}\ \bibnamefont {Noya}},\ }\href@noop {}
  {\bibfield  {journal} {\bibinfo  {journal} {J. Phys.: Condens. Matter}\
  }\textbf {\bibinfo {volume} {20}},\ \bibinfo {pages} {153101} (\bibinfo
  {year} {2008})}\BibitemShut {NoStop}%
\bibitem [{\citenamefont {Wilding}\ and\ \citenamefont
  {Bruce}(2000)}]{wilding:00}%
  \BibitemOpen
  \bibfield  {author} {\bibinfo {author} {\bibfnamefont {N.~B.}\ \bibnamefont
  {Wilding}}\ and\ \bibinfo {author} {\bibfnamefont {A.~D.}\ \bibnamefont
  {Bruce}},\ }\href@noop {} {\bibfield  {journal} {\bibinfo  {journal} {Phys.
  Rev. Lett.}\ }\textbf {\bibinfo {volume} {85}},\ \bibinfo {pages} {5138}
  (\bibinfo {year} {2000})}\BibitemShut {NoStop}%
\bibitem [{\citenamefont {Errington}(2004)}]{errington:04}%
  \BibitemOpen
  \bibfield  {author} {\bibinfo {author} {\bibfnamefont {J.~R.}\ \bibnamefont
  {Errington}},\ }\href@noop {} {\bibfield  {journal} {\bibinfo  {journal} {J.
  Chem. Phys.}\ }\textbf {\bibinfo {volume} {120}},\ \bibinfo {pages} {3130}
  (\bibinfo {year} {2004})}\BibitemShut {NoStop}%
\bibitem [{\citenamefont {Hoover}\ and\ \citenamefont {Ree}(1968)}]{hoover:68}%
  \BibitemOpen
  \bibfield  {author} {\bibinfo {author} {\bibfnamefont {W.~G.}\ \bibnamefont
  {Hoover}}\ and\ \bibinfo {author} {\bibfnamefont {F.~H.}\ \bibnamefont
  {Ree}},\ }\href@noop {} {\bibfield  {journal} {\bibinfo  {journal} {J. Chem.
  Phys}\ }\textbf {\bibinfo {volume} {49}},\ \bibinfo {pages} {3609} (\bibinfo
  {year} {1968})}\BibitemShut {NoStop}%
\bibitem [{\citenamefont {Frenkel}\ and\ \citenamefont
  {Ladd}(1984)}]{frenkel:84}%
  \BibitemOpen
  \bibfield  {author} {\bibinfo {author} {\bibfnamefont {D.}~\bibnamefont
  {Frenkel}}\ and\ \bibinfo {author} {\bibfnamefont {A.~K.~C.}\ \bibnamefont
  {Ladd}},\ }\href@noop {} {\bibfield  {journal} {\bibinfo  {journal} {J. Chem.
  Phys.}\ }\textbf {\bibinfo {volume} {81}},\ \bibinfo {pages} {3188} (\bibinfo
  {year} {1984})}\BibitemShut {NoStop}%
\bibitem [{\citenamefont {Polson}\ \emph {et~al.}(2000)\citenamefont {Polson},
  \citenamefont {Trizac}, \citenamefont {Pronk},\ and\ \citenamefont
  {Frenkel}}]{polson:00}%
  \BibitemOpen
  \bibfield  {author} {\bibinfo {author} {\bibfnamefont {J.~M.}\ \bibnamefont
  {Polson}}, \bibinfo {author} {\bibfnamefont {E.}~\bibnamefont {Trizac}},
  \bibinfo {author} {\bibfnamefont {S.}~\bibnamefont {Pronk}}, \ and\ \bibinfo
  {author} {\bibfnamefont {D.}~\bibnamefont {Frenkel}},\ }\href@noop {}
  {\bibfield  {journal} {\bibinfo  {journal} {J. Chem. Phys.}\ }\textbf
  {\bibinfo {volume} {112}},\ \bibinfo {pages} {5339} (\bibinfo {year}
  {2000})}\BibitemShut {NoStop}%
\bibitem [{\citenamefont {Vega}\ and\ \citenamefont {Noya}(2007)}]{vega:07}%
  \BibitemOpen
  \bibfield  {author} {\bibinfo {author} {\bibfnamefont {C.}~\bibnamefont
  {Vega}}\ and\ \bibinfo {author} {\bibfnamefont {E.~G.}\ \bibnamefont
  {Noya}},\ }\href@noop {} {\bibfield  {journal} {\bibinfo  {journal} {J. Chem.
  Phys.}\ }\textbf {\bibinfo {volume} {127}},\ \bibinfo {pages} {154113}
  (\bibinfo {year} {2007})}\BibitemShut {NoStop}%
\bibitem [{\citenamefont {Ladd}\ and\ \citenamefont
  {Woodcock}(1977)}]{ladd:77}%
  \BibitemOpen
  \bibfield  {author} {\bibinfo {author} {\bibfnamefont {J.~C.}\ \bibnamefont
  {Ladd}}\ and\ \bibinfo {author} {\bibfnamefont {L.~V.}\ \bibnamefont
  {Woodcock}},\ }\href@noop {} {\bibfield  {journal} {\bibinfo  {journal} {J.
  Chem. Phys.}\ }\textbf {\bibinfo {volume} {51}},\ \bibinfo {pages} {155}
  (\bibinfo {year} {1977})}\BibitemShut {NoStop}%
\bibitem [{\citenamefont {Noya}\ \emph {et~al.}(2008)\citenamefont {Noya},
  \citenamefont {Vega},\ and\ \citenamefont {de~Miguel}}]{noya:08}%
  \BibitemOpen
  \bibfield  {author} {\bibinfo {author} {\bibfnamefont {E.~G.}\ \bibnamefont
  {Noya}}, \bibinfo {author} {\bibfnamefont {C.}~\bibnamefont {Vega}}, \ and\
  \bibinfo {author} {\bibfnamefont {E.}~\bibnamefont {de~Miguel}},\ }\href@noop
  {} {\bibfield  {journal} {\bibinfo  {journal} {J. Chem. Phys.}\ }\textbf
  {\bibinfo {volume} {128}},\ \bibinfo {pages} {154507} (\bibinfo {year}
  {2008})}\BibitemShut {NoStop}%
\bibitem [{\citenamefont {Zykova-Timan}\ \emph {et~al.}(2010)\citenamefont
  {Zykova-Timan}, \citenamefont {Horbach},\ and\ \citenamefont
  {Binder}}]{zykova-timan:10}%
  \BibitemOpen
  \bibfield  {author} {\bibinfo {author} {\bibfnamefont {T.}~\bibnamefont
  {Zykova-Timan}}, \bibinfo {author} {\bibfnamefont {J.}~\bibnamefont
  {Horbach}}, \ and\ \bibinfo {author} {\bibfnamefont {K.}~\bibnamefont
  {Binder}},\ }\href@noop {} {\bibfield  {journal} {\bibinfo  {journal} {J.
  Chem. Phys}\ }\textbf {\bibinfo {volume} {133}},\ \bibinfo {pages} {014705}
  (\bibinfo {year} {2010})}\BibitemShut {NoStop}%
\bibitem [{\citenamefont {Davidchack}(2010)}]{davidchack:10}%
  \BibitemOpen
  \bibfield  {author} {\bibinfo {author} {\bibfnamefont {R.~L.}\ \bibnamefont
  {Davidchack}},\ }\href@noop {} {\bibfield  {journal} {\bibinfo  {journal} {J.
  Chem. Phys.}\ }\textbf {\bibinfo {volume} {133}},\ \bibinfo {pages} {234701}
  (\bibinfo {year} {2010})}\BibitemShut {NoStop}%
\bibitem [{\citenamefont {Cacciuto}\ \emph {et~al.}(2003)\citenamefont
  {Cacciuto}, \citenamefont {Auer},\ and\ \citenamefont
  {Frenkel}}]{cacciuto:03}%
  \BibitemOpen
  \bibfield  {author} {\bibinfo {author} {\bibfnamefont {A.}~\bibnamefont
  {Cacciuto}}, \bibinfo {author} {\bibfnamefont {S.}~\bibnamefont {Auer}}, \
  and\ \bibinfo {author} {\bibfnamefont {D.}~\bibnamefont {Frenkel}},\
  }\href@noop {} {\bibfield  {journal} {\bibinfo  {journal} {J. Chem. Phys.}\
  }\textbf {\bibinfo {volume} {119}},\ \bibinfo {pages} {7467} (\bibinfo {year}
  {2003})}\BibitemShut {NoStop}%
\bibitem [{\citenamefont {Mu}\ \emph {et~al.}(2005)\citenamefont {Mu},
  \citenamefont {Houk},\ and\ \citenamefont {Song}}]{mu:05}%
  \BibitemOpen
  \bibfield  {author} {\bibinfo {author} {\bibfnamefont {Y.}~\bibnamefont
  {Mu}}, \bibinfo {author} {\bibfnamefont {A.}~\bibnamefont {Houk}}, \ and\
  \bibinfo {author} {\bibfnamefont {X.}~\bibnamefont {Song}},\ }\href@noop {}
  {\bibfield  {journal} {\bibinfo  {journal} {J. Phys. Chem. B}\ }\textbf
  {\bibinfo {volume} {109}},\ \bibinfo {pages} {6500} (\bibinfo {year}
  {2005})}\BibitemShut {NoStop}%
\bibitem [{\citenamefont {Fernandez}\ \emph {et~al.}(2009)\citenamefont
  {Fernandez}, \citenamefont {Martin-Mayor},\ and\ \citenamefont
  {Yllanes}}]{fernandez:09}%
  \BibitemOpen
  \bibfield  {author} {\bibinfo {author} {\bibfnamefont {L.~A.}\ \bibnamefont
  {Fernandez}}, \bibinfo {author} {\bibfnamefont {V.}~\bibnamefont
  {Martin-Mayor}}, \ and\ \bibinfo {author} {\bibfnamefont {D.}~\bibnamefont
  {Yllanes}},\ }\href@noop {} {\bibfield  {journal} {\bibinfo  {journal} {Nucl.
  Phys. B}\ }\textbf {\bibinfo {volume} {807}},\ \bibinfo {pages} {424}
  (\bibinfo {year} {2009})}\BibitemShut {NoStop}%
\bibitem [{\citenamefont {Martin-Mayor}\ \emph {et~al.}(2011)\citenamefont
  {Martin-Mayor}, \citenamefont {Seoane},\ and\ \citenamefont
  {Yllanes}}]{martin-mayor:11}%
  \BibitemOpen
  \bibfield  {author} {\bibinfo {author} {\bibfnamefont {V.}~\bibnamefont
  {Martin-Mayor}}, \bibinfo {author} {\bibfnamefont {B.}~\bibnamefont
  {Seoane}}, \ and\ \bibinfo {author} {\bibfnamefont {D.}~\bibnamefont
  {Yllanes}},\ }\href {\doibase doi:10.1007/s10955-011-0261-4} {\bibfield
  {journal} {\bibinfo  {journal} {J. Stat. Phys.}\ }\textbf {\bibinfo {volume}
  {144}},\ \bibinfo {pages} {554} (\bibinfo {year} {2011})}\BibitemShut
  {NoStop}%
\bibitem [{\citenamefont {Torrie}\ and\ \citenamefont
  {Valleau}(1974)}]{torrie:74}%
  \BibitemOpen
  \bibfield  {author} {\bibinfo {author} {\bibfnamefont {G.~M.}\ \bibnamefont
  {Torrie}}\ and\ \bibinfo {author} {\bibfnamefont {J.~P.}\ \bibnamefont
  {Valleau}},\ }\href@noop {} {\bibfield  {journal} {\bibinfo  {journal} {Chem.
  Phys. Lett.}\ }\textbf {\bibinfo {volume} {28}},\ \bibinfo {pages} {578}
  (\bibinfo {year} {1974})}\BibitemShut {NoStop}%
\bibitem [{\citenamefont {Torrie}\ and\ \citenamefont
  {Valleau}(1977)}]{torrie:77}%
  \BibitemOpen
  \bibfield  {author} {\bibinfo {author} {\bibfnamefont {G.~M.}\ \bibnamefont
  {Torrie}}\ and\ \bibinfo {author} {\bibfnamefont {J.~P.}\ \bibnamefont
  {Valleau}},\ }\href@noop {} {\bibfield  {journal} {\bibinfo  {journal} {J.
  Comp. Physics.}\ }\textbf {\bibinfo {volume} {23}},\ \bibinfo {pages} {187}
  (\bibinfo {year} {1977})}\BibitemShut {NoStop}%
\bibitem [{\citenamefont {Bartels}(2000)}]{bartels:00}%
  \BibitemOpen
  \bibfield  {author} {\bibinfo {author} {\bibfnamefont {C.}~\bibnamefont
  {Bartels}},\ }\href@noop {} {\bibfield  {journal} {\bibinfo  {journal} {Chem.
  Phys. Lett.}\ }\textbf {\bibinfo {volume} {331}},\ \bibinfo {pages} {446}
  (\bibinfo {year} {2000})}\BibitemShut {NoStop}%
\bibitem [{\citenamefont {ten Wolde}\ \emph {et~al.}(1995)\citenamefont {ten
  Wolde}, \citenamefont {Ruiz-Montero},\ and\ \citenamefont
  {Frenkel}}]{tenwolde:95}%
  \BibitemOpen
  \bibfield  {author} {\bibinfo {author} {\bibfnamefont {P.}~\bibnamefont {ten
  Wolde}}, \bibinfo {author} {\bibfnamefont {M.~J.}\ \bibnamefont
  {Ruiz-Montero}}, \ and\ \bibinfo {author} {\bibfnamefont {D.}~\bibnamefont
  {Frenkel}},\ }\href@noop {} {\bibfield  {journal} {\bibinfo  {journal} {Phys.
  Rev. Lett.}\ }\textbf {\bibinfo {volume} {75}},\ \bibinfo {pages} {2714}
  (\bibinfo {year} {1995})}\BibitemShut {NoStop}%
\bibitem [{\citenamefont {Schrader}\ \emph {et~al.}(2009)\citenamefont
  {Schrader}, \citenamefont {Virnau},\ and\ \citenamefont
  {Binder}}]{schrader:09}%
  \BibitemOpen
  \bibfield  {author} {\bibinfo {author} {\bibfnamefont {M.}~\bibnamefont
  {Schrader}}, \bibinfo {author} {\bibfnamefont {P.}~\bibnamefont {Virnau}}, \
  and\ \bibinfo {author} {\bibfnamefont {K.}~\bibnamefont {Binder}},\
  }\href@noop {} {\bibfield  {journal} {\bibinfo  {journal} {Phys. Rev. E}\
  }\textbf {\bibinfo {volume} {79}},\ \bibinfo {pages} {061104} (\bibinfo
  {year} {2009})}\BibitemShut {NoStop}%
\bibitem [{\citenamefont {Binder}\ \emph {et~al.}(2011)\citenamefont {Binder},
  \citenamefont {Block}, \citenamefont {Das}, \citenamefont {Virnau},\ and\
  \citenamefont {Winter}}]{binder:11}%
  \BibitemOpen
  \bibfield  {author} {\bibinfo {author} {\bibfnamefont {K.}~\bibnamefont
  {Binder}}, \bibinfo {author} {\bibfnamefont {B.}~\bibnamefont {Block}},
  \bibinfo {author} {\bibfnamefont {S.~K.}\ \bibnamefont {Das}}, \bibinfo
  {author} {\bibfnamefont {P.}~\bibnamefont {Virnau}}, \ and\ \bibinfo {author}
  {\bibfnamefont {D.}~\bibnamefont {Winter}},\ }\href@noop {} {\  (\bibinfo
  {year} {2011})},\ \Eprint {http://arxiv.org/abs/arXiv:1103.2241}
  {arXiv:1103.2241} \BibitemShut {NoStop}%
\bibitem [{\citenamefont {Steinhardt}\ \emph {et~al.}(1983)\citenamefont
  {Steinhardt}, \citenamefont {Nelson},\ and\ \citenamefont
  {Ronchetti}}]{steinhardt:83}%
  \BibitemOpen
  \bibfield  {author} {\bibinfo {author} {\bibfnamefont {P.~J.}\ \bibnamefont
  {Steinhardt}}, \bibinfo {author} {\bibfnamefont {D.~R.}\ \bibnamefont
  {Nelson}}, \ and\ \bibinfo {author} {\bibfnamefont {M.}~\bibnamefont
  {Ronchetti}},\ }\href@noop {} {\bibfield  {journal} {\bibinfo  {journal}
  {Phys. Rev. B}\ }\textbf {\bibinfo {volume} {28}},\ \bibinfo {pages} {784}
  (\bibinfo {year} {1983})}\BibitemShut {NoStop}%
\bibitem [{\citenamefont {van Duijneveldt}\ and\ \citenamefont
  {Frenkel}(1992)}]{duijneveldt:92}%
  \BibitemOpen
  \bibfield  {author} {\bibinfo {author} {\bibfnamefont {J.~S.}\ \bibnamefont
  {van Duijneveldt}}\ and\ \bibinfo {author} {\bibfnamefont {D.}~\bibnamefont
  {Frenkel}},\ }\href@noop {} {\bibfield  {journal} {\bibinfo  {journal} {J.
  Chem. Phys}\ }\textbf {\bibinfo {volume} {96}},\ \bibinfo {pages} {4655}
  (\bibinfo {year} {1992})}\BibitemShut {NoStop}%
\bibitem [{\citenamefont {Angioletti-Uberti}\ \emph {et~al.}(2010)\citenamefont
  {Angioletti-Uberti}, \citenamefont {Ceriotti}, \citenamefont {Lee},\ and\
  \citenamefont {Finnis}}]{angioletti:10}%
  \BibitemOpen
  \bibfield  {author} {\bibinfo {author} {\bibfnamefont {S.}~\bibnamefont
  {Angioletti-Uberti}}, \bibinfo {author} {\bibfnamefont {M.}~\bibnamefont
  {Ceriotti}}, \bibinfo {author} {\bibfnamefont {P.~D.}\ \bibnamefont {Lee}}, \
  and\ \bibinfo {author} {\bibfnamefont {M.~W.}\ \bibnamefont {Finnis}},\
  }\href@noop {} {\bibfield  {journal} {\bibinfo  {journal} {Phys. Rev. B}\
  }\textbf {\bibinfo {volume} {81}},\ \bibinfo {pages} {125416} (\bibinfo
  {year} {2010})}\BibitemShut {NoStop}%
\bibitem [{\citenamefont {Biskup}\ \emph {et~al.}(2002)\citenamefont {Biskup},
  \citenamefont {Chayes},\ and\ \citenamefont {Koteck{\'y}}}]{biskup:02}%
  \BibitemOpen
  \bibfield  {author} {\bibinfo {author} {\bibfnamefont {M.}~\bibnamefont
  {Biskup}}, \bibinfo {author} {\bibfnamefont {L.}~\bibnamefont {Chayes}}, \
  and\ \bibinfo {author} {\bibfnamefont {R.}~\bibnamefont {Koteck{\'y}}},\
  }\href@noop {} {\bibfield  {journal} {\bibinfo  {journal} {Europhys. Lett.}\
  }\textbf {\bibinfo {volume} {60}},\ \bibinfo {pages} {21} (\bibinfo {year}
  {2002})}\BibitemShut {NoStop}%
\bibitem [{\citenamefont {Binder}(2003)}]{binder:03}%
  \BibitemOpen
  \bibfield  {author} {\bibinfo {author} {\bibfnamefont {K.}~\bibnamefont
  {Binder}},\ }\href@noop {} {\bibfield  {journal} {\bibinfo  {journal}
  {Physica A}\ }\textbf {\bibinfo {volume} {319}},\ \bibinfo {pages} {99}
  (\bibinfo {year} {2003})}\BibitemShut {NoStop}%
\bibitem [{\citenamefont {MacDowell}\ \emph {et~al.}(2006)\citenamefont
  {MacDowell}, \citenamefont {Shen},\ and\ \citenamefont
  {Errington}}]{macdowell:06}%
  \BibitemOpen
  \bibfield  {author} {\bibinfo {author} {\bibfnamefont {L.~G.}\ \bibnamefont
  {MacDowell}}, \bibinfo {author} {\bibfnamefont {V.}~\bibnamefont {Shen}}, \
  and\ \bibinfo {author} {\bibfnamefont {J.~R.}\ \bibnamefont {Errington}},\
  }\href@noop {} {\bibfield  {journal} {\bibinfo  {journal} {J. Chem. Phys.}\
  }\textbf {\bibinfo {volume} {125}},\ \bibinfo {pages} {034705} (\bibinfo
  {year} {2006})}\BibitemShut {NoStop}%
\bibitem [{\citenamefont {Bolhuis}\ \emph {et~al.}(1997)\citenamefont
  {Bolhuis}, \citenamefont {Frenkel}, \citenamefont {Mau},\ and\ \citenamefont
  {Huse}}]{bolhuis:97}%
  \BibitemOpen
  \bibfield  {author} {\bibinfo {author} {\bibfnamefont {P.-G.}\ \bibnamefont
  {Bolhuis}}, \bibinfo {author} {\bibfnamefont {D.}~\bibnamefont {Frenkel}},
  \bibinfo {author} {\bibfnamefont {S.~C.}\ \bibnamefont {Mau}}, \ and\
  \bibinfo {author} {\bibfnamefont {D.~A.}\ \bibnamefont {Huse}},\ }\href@noop
  {} {\bibfield  {journal} {\bibinfo  {journal} {Nature}\ }\textbf {\bibinfo
  {volume} {388}},\ \bibinfo {pages} {235} (\bibinfo {year}
  {1997})}\BibitemShut {NoStop}%
\bibitem [{Note1()}]{Note1}%
  \BibitemOpen
  \bibinfo {note} {Particles $i$ and $j$ are \protect \textit {neighbors} if
  $r_{ij}<1.5\ \sigma $. In the ideal FCC structure, for all particle-densities
  relevant to us, this choice includes only the nearest neighbors
  shell.}\BibitemShut {Stop}%
\bibitem [{Note2()}]{Note2}%
  \BibitemOpen
  \bibinfo {note} {Eq.~\protect \textup {\hbox {\mathsurround \z@ \protect
  \normalfont (\ignorespaces \ref {eq:weight}\unskip \@@italiccorr )}} behaves
  as an animal's tether: only if (say) $|{\protect \mathaccentV
  {hat}05EQ}_6-Q_6(\protect \boldsymbol {R})|\gg 1/\protect \sqrt {N\alpha }$
  is the penalty large. Note as well that Eqs.~(\ref {eq:weight},\ref {eq:h})
  generalize straightforwardly to the case of more than two
  quasi-constraints.}\BibitemShut {Stop}%
\bibitem [{Note3()}]{Note3}%
  \BibitemOpen
  \bibinfo {note} {A magnitude $A$ is {\protect \em additive} if $N A$ is
  extensive: gluing together systems 1,2 (with $N^{(i)}$ particles and
  $A=A^{(i)}$, $i=1,2$), results in a total system with $N=N^{(1)}+N^{(2)}$
  particles and $NA=N^{(1)} A^{(1)} +N^{(2)}A^{(2)}$ (plus subdominant
  corrections such as surface effects $\sim N^{2/3}$). $C$ is additive to a
  great accuracy for coexisting fluid and FCC phases, because the average
  number of neighbors $N_b$ is very similar in both phases ($5\%$ difference,
  with negligible effects on additivity in our $N$ range, as compared with
  surface effects). $Q_6$ is additive only if one of the subsystems, say $i=1$,
  is a liquid so that $Q_6^{(1)}\sim 1/\protect \sqrt {N^{(1)}}$ ($Q_6$ is a
  pseudo-order parameter, i.e. a strictly positive quantity which is of order
  $1/\protect \sqrt {N}$ in a disordered phase). For studies of interfaces on
  larger systems, it would be advisable to choose exactly additive order
  parameters.}\BibitemShut {Stop}%
\bibitem [{\citenamefont {Ruelle}(1969)}]{ruelle:69}%
  \BibitemOpen
  \bibfield  {author} {\bibinfo {author} {\bibfnamefont {D.}~\bibnamefont
  {Ruelle}},\ }\href@noop {} {\emph {\bibinfo {title} {Statistical
  Mechanics}}}\ (\bibinfo  {publisher} {Benjamin},\ \bibinfo {year}
  {1969})\BibitemShut {NoStop}%
\bibitem [{Note4()}]{Note4}%
  \BibitemOpen
  \bibinfo {note} {Our runs for $N\leq 2916$ are, at least, $100 \tau $ long
  ($\tau $ is the integrated autocorrelation time~\cite {sokal:97}, computed
  for $Q_6$ and $v$~\cite {martin-mayor:11}). For $N=2916$, but only at
  $S=0.4$, we find metastability with a helicoidal configuration (however, its
  contribution to final quantities is smaller than statistical errors).
  Metastabilities arise often for $N=4000$, at intermediate $S$ (yet, a careful
  selection of starting configurations yields a $\protect \boldsymbol {\nabla
  }\varOmega _N$ with smooth $S$ dependency).}\BibitemShut {Stop}%
\bibitem [{\citenamefont {Ferrenberg}\ and\ \citenamefont
  {Swendsen}(1988)}]{ferrenberg:88}%
  \BibitemOpen
  \bibfield  {author} {\bibinfo {author} {\bibfnamefont {A.~M.}\ \bibnamefont
  {Ferrenberg}}\ and\ \bibinfo {author} {\bibfnamefont {R.~H.}\ \bibnamefont
  {Swendsen}},\ }\href@noop {} {\bibfield  {journal} {\bibinfo  {journal}
  {Phys. Rev. Lett.}\ }\textbf {\bibinfo {volume} {61}},\ \bibinfo {pages}
  {2635} (\bibinfo {year} {1988})}\BibitemShut {NoStop}%
\bibitem [{\citenamefont {Borgs}\ and\ \citenamefont
  {Koteck\'y}(1992)}]{borgs:92}%
  \BibitemOpen
  \bibfield  {author} {\bibinfo {author} {\bibfnamefont {C.}~\bibnamefont
  {Borgs}}\ and\ \bibinfo {author} {\bibfnamefont {R.}~\bibnamefont
  {Koteck\'y}},\ }\href@noop {} {\bibfield  {journal} {\bibinfo  {journal}
  {Phys. Rev. Lett.}\ }\textbf {\bibinfo {volume} {68}},\ \bibinfo {pages}
  {1734} (\bibinfo {year} {1992})}\BibitemShut {NoStop}%
\bibitem [{Note5()}]{Note5}%
  \BibitemOpen
  \bibinfo {note} {The tethering approach should not induce artificial 
  interfaces. In fact, mathematically, the interfacial free-energy is defined
  though the ratio of two partition functions with different boundary
  conditions. The tethered potential does not change the partition function
  [with any boundary conditions, see Eq.~\protect \textup {\hbox {\mathsurround
  \z@ \protect \normalfont (\ignorespaces \ref
  {eq:ensemble-independence}\unskip \@@italiccorr )}}].}\BibitemShut {Stop}%
\bibitem [{\citenamefont {Binder}(1982)}]{binder:82}%
  \BibitemOpen
  \bibfield  {author} {\bibinfo {author} {\bibfnamefont {K.}~\bibnamefont
  {Binder}},\ }\href@noop {} {\bibfield  {journal} {\bibinfo  {journal} {Phys.
  Rev. A}\ }\textbf {\bibinfo {volume} {25}},\ \bibinfo {pages} {1699}
  (\bibinfo {year} {1982})}\BibitemShut {NoStop}%
\bibitem [{\citenamefont {Billoire}\ \emph {et~al.}(1994)\citenamefont
  {Billoire}, \citenamefont {Neuhaus},\ and\ \citenamefont
  {Berg}}]{billoire:94}%
  \BibitemOpen
  \bibfield  {author} {\bibinfo {author} {\bibfnamefont {A.}~\bibnamefont
  {Billoire}}, \bibinfo {author} {\bibfnamefont {T.}~\bibnamefont {Neuhaus}}, \
  and\ \bibinfo {author} {\bibfnamefont {B.}~\bibnamefont {Berg}},\ }\href@noop
  {} {\bibfield  {journal} {\bibinfo  {journal} {Nucl. Phys. B}\ }\textbf
  {\bibinfo {volume} {413}},\ \bibinfo {pages} {795} (\bibinfo {year}
  {1994})}\BibitemShut {NoStop}%
\bibitem [{\citenamefont {Sokal}(1997)}]{sokal:97}%
  \BibitemOpen
  \bibfield  {author} {\bibinfo {author} {\bibfnamefont {A.~D.}\ \bibnamefont
  {Sokal}},\ }in\ \href@noop {} {\emph {\bibinfo {booktitle} {Functional
  Integration: Basics and Applications (1996 Carg{\`e}se School)}}},\ \bibinfo
  {editor} {edited by\ \bibinfo {editor} {\bibfnamefont {C.}~\bibnamefont
  {DeWitt-Morette}}, \bibinfo {editor} {\bibfnamefont {P.}~\bibnamefont
  {Cartier}}, \ and\ \bibinfo {editor} {\bibfnamefont {A.}~\bibnamefont
  {Folacci}}}\ (\bibinfo  {publisher} {Plenum},\ \bibinfo {address} {N.Y.},\
  \bibinfo {year} {1997})\BibitemShut {NoStop}%
\end{thebibliography}

%

\end{document}